# Cost-Efficient Orchestration of Containers in Clouds: A Vision, Architectural Elements, and Future Directions

**Rajkumar Buyya[1], Maria A. Rodriguez[1], Adel Nadjaran Toosi[2], Jaeman Park[3]**

[1]Cloud Computing and Distributed Systems (CLOUDS) Lab
School of Computing and Information Systems
The University of Melbourne, Australia

[2]Faculty of Information Technology
Monash University, Australia

[3]Data Cloud Lab, AI Center, Samsung Research
Samsung Electronics, South Korea

**Abstract.** This paper proposes an architectural framework for the efficient orchestration of containers in cloud environments. It centres around resource scheduling and rescheduling policies as well as autoscaling algorithms that enable the creation of elastic virtual clusters. In this way, the proposed framework enables the sharing of a computing environment between differing client applications packaged in containers, including web services, offline analytics jobs, and backend pre-processing tasks. The devised resource management algorithms and policies will improve utilization of the available virtual resources to reduce operational cost for the provider while satisfying the resource needs of various types of applications. The proposed algorithms will take factors that are previously omitted by other solutions into consideration, including 1) the pricing models of the acquired resources, 2) and the fault-tolerability of the applications, and 3) the QoS requirements of the running applications, such as the latencies and throughputs of the web services and the deadline of the analytical and pre-processing jobs. The proposed solutions will be evaluated by developing a prototype platform based on one of the existing container orchestration platforms.

## 1. Introduction

Containers, enabling lightweight environment and performance isolation, fast and flexible deployment, and fine-grained resource sharing, have gained popularity in better management and application deployment in addition to hardware virtualization [19][20]. Recently, various container solutions, such as Docker [1] and LXC [2], have gained acceptance by organizations to run various kinds of applications on either proprietary data centers or virtual clusters in Clouds. To efficiently utilize resources, instead of running separate clusters of homogeneous containers, organizations prefer to run containers of different applications on a shared cluster, which drives the creation of modern container orchestration platforms, such as Kubernetes [3], Docker Swarm [4], and Apache Mesos [5]. These platforms provide efficient bin-packing algorithms to schedule containers on the shared clusters. However, their solutions are static and cannot meet the dynamic resource needs of the applications by failing to consider the characteristics of the Cloud environments and applications.

In the context of Cloud, container orchestration platforms are usually deployed on virtual clusters acquired from a single or multiple Cloud data centers. To exploit the elasticity feature of Cloud, they support dynamically adding or removing resources from the cluster. Therefore, organizations can utilize the autoscaling systems provided by Cloud providers or their own solutions to automatically adjust the size of the shared clusters according to their real-time utilization. To save operational cost in Cloud, it is essential to consolidate containers onto as fewer virtual machines as possible because the initial placement may deteriorate along with dynamisms caused by workload fluctuation, application launches and terminations, and pricing variations. To realize this goal, the platform should be able to migrate the containers running on underutilized VMs to other VMs. Using the state-of-the-art techniques, container migration can be implemented in three steps: 1) checkpointing the target container (if necessary) [6], 2) killing it on the original host, and 3) then resuming it on the target host, which is only applicable to stateless and fault-tolerant applications. Therefore, the initial container placement strategy is as important as the dynamic rescheduling algorithm to the success of a container management platform in Cloud.

Currently, Cloud providers offer three types of pricing mechanisms, including reserved, on-demand, and rebated pre-emptive mode. In order to balance cost-efficiency, reliability, and QoS, organizations can provision their clusters dynamically using mixed instances of the three pricing models according to real-time workloads. This requires the orchestration platforms to be aware of the pricing models of the underlying resources, and the characteristics (e.g., ability to tolerate faults/failures) of the applications in order to correctly and wisely schedule and reschedule them on appropriate resources. For example, the orchestration algorithm should never place customer-facing applications on rebated preemptive resources since user experience will be disrupted by preemptions of the VMs. In addition, long-running jobs like web services should be avoided to be deployed on on-demand resources as it will prolong the running time of expensive on-demand resources. On the other hand, fault-tolerant applications are welcomed to run on rebated preemptive resources and on-demand resources as they can tolerate resource failures and can be terminated anytime to free the expensive on-demand resources.

Taking all these factors and requirements into account, the container orchestration problem becomes a challenging task. In summary, it involves three sub-problems: 1) efficiently schedule the containers onto suitable resources in their initial placement attempts so that later migrations are minimized and VMs can be freed easily; 2) dynamically reschedule movable containers to free up more resources while satisfying QoS requirements of the applications; 3) dynamically provision resources of different pricing models to the virtual cluster in order to satisfy growing needs of the applications with minimum cost.

*1.1. Research Methodology*

The methodology for solving the problem of scheduling, rescheduling and scaling policies for shared container-based virtual clusters in Clouds so that resource utilization is maximized, and resource cost is minimized while meeting QoS requirements of various applications running on the clusters is as follows:
- Define an architectural framework and principles for QoS-aware cluster management for containers.
- Investigate and develop efficient scheduling and rescheduling algorithms and scaling policies for container-based clusters in Clouds.
- Develop techniques that overcome the complexity caused by heterogeneities in pricing, application fault-tolerability, and QoS requirements, and exploit those heterogeneities to improve resource utilization and reduce cost.
- Develop a prototype system incorporating the techniques proposed above.

The rest of this paper is organized as follows: First, a brief survey on the existing literature in container orchestration systems is presented. Next, we present the proposed system architecture and explain its elements in detail, followed by the prototype framework implementation and some preliminary experiments and results to show the potential effectiveness of the proposed approaches. Finally, the paper concludes with a discussion on approaches for realizing future directions.

## 2. Related Work

There have been considerable attempts to build a reliable and scalable container management platform for clusters from the research and open-source communities. These systems target to ease the deployment and scaling of general-purpose applications packaged using container technologies such as Docker [1] and LXC [2] containers. Many of the solutions have been developed and deployed in very large-scale environments for production purposes by leading companies in various industries either on proprietary infrastructures or on public Clouds.

Kubernetes [3], derived from Google's in-house container management platform Borg [7], is an open-source system that enables automatic deployment and scaling of containerized applications. In addition, it also offers self-healing capacities that kill and replace unresponsive nodes and containers, and automated rollout and rollback features for applications and their configurations. The scheduler of Kubernetes is pluggable, and it provides a default scheduler that in turn schedules each container group (known as pods) on available resources filtered by user-defined requirements and ranked based on individually defined application affinities.

Docker Swarm [4] is the native clustering solution for Docker containers. It provides discovery services and schedules containers onto hosts using pre-defined strategies and filters. By default, it supports two simple strategies: (1) spread strategy, which favors running new containers on least loaded hosts, and (2) BinPack strategy, which selects the most loaded host that have enough resources to run the containers. Docker Swarm also employs a filtering mechanism to map containers to resources. Users can define filters regarding host statuses (e.g., based on available CPU and memory resources and health check results) and container configurations (e.g., resource and file affinities and dependencies to other containers).

Nomad [8] is an enterprise cluster management product from HashiCorp. It not only can deploy containers but also virtual machines and application runtimes like JVM. In addition, it is also capable of operating across multiple data centers and providing high-availability.

Apache Mesos [5] enables cluster sharing among various applications. Different from the previous platforms, Mesos can be viewed as a meta-platform that operates above them. It employs a two-stage scheduling approach. In the first stage, it divides the resources of the cluster and respectively offers available resources to each application, called framework. The framework once accepts the offer can then schedule its tasks on the obtained resources using its own framework scheduler. After that, Mesos actually launches the tasks for the framework on the corresponding hosts. Kubernetes [3] has been integrated into Mesos as a framework to run general-purpose containers. Apache Marathon [9] and Apache Aurora [10] are two other popular general-purpose frameworks built on Mesos. In addition, common data analytics platforms, such as Apache Hadoop [11], Apache Spark [12], and Apache Storm [13], and distributed job schedulers, such as Dkron [14] and Chronos [15], all support running on Mesos. The unique feature of its two-phase scheduling enables Mesos to run frameworks based on their urgency and priority while easing dynamic resource provisioning.

The scheduling and scaling modules of the existing platforms noted above only provide rudimentary strategies that overlook the heterogeneities in pricing models of Cloud resources, fault-tolerability of applications, and QoS requirements of applications. It is essential to take these factors into account to further improve resource utilization and reduce the cost of virtual clusters in public Clouds.

## 3. Significance

Along with the development and application of new technologies, including IoT, and artificial intelligence, the need for computing power will continue to grow significantly in the future. Cloud computing which has become the backbone of the IT infrastructure is expected to play a major role to satisfy this incoming computing demands. Therefore, minimizing the expenses of leasing Cloud resources becomes ever increasingly important for organizations relying on Clouds to host their applications and support their core business operation.

In addition, with the container management platforms enabling easy and flexible deployment and resource sharing on the same virtual cluster, organizations have been attracted to provision their computing infrastructure in this manner instead of managing separate virtual clusters for different applications.

Our research comes into the center of helping organizations to reduce their Cloud resource leasing expenses by utilizing this new resource management paradigm to improve their profitability and sustainability of businesses. This paper identifies novel resource scheduling/rescheduling and scaling policies that comprehensively consider the pricing mechanisms of Cloud resources, and fault-tolerability of applications to further reduce the cost of running a containerized cluster in Clouds while satisfying the individual QoS requirements of the applications sharing the platform.

## 4. System Architecture

The paper explores the problem of pricing, fault-tolerability, and QoS-aware scheduling/rescheduling algorithms and scaling policies for containerized clusters in Clouds. An architectural framework for the realization of these goals is shown in Figure 1. There are three main entities involved in the illustrated architecture; the users and applications, the containerized cluster management system, and the cloud resources.

*Consumers and Applications:* Cloud consumers acquire virtual or physical resources from Cloud data centers located anywhere in the world and compose a shared cluster using container management software. They submit their applications packaged in containers to the cluster to execute. The concept of consumers here refers to the organizations that rely on Clouds as the computing platform to deliver operations. They themselves may target to serve the end "users" submitting requests from terminal devices. An application refers to a set of one or more jobs packaged in one or more containers. Various applications are deployed on a shared platform or cluster. These can be user-facing web services, internal backend real-time services, data analytics jobs, data pre-processing tasks, cron jobs, etc. The cloud consumer should specify some meta-data for each application, such as the detailed QoS requirements in measurable metrics and whether the framework can tolerate failures or not.

*Containerized Cluster Manager:* Acts as the interface between the obtained virtual/physical resources and application frameworks and manager of the running applications. It requires the interaction of the following components to support cost-efficient and QoS-aware resource management:

<u>Scheduler:</u> Allocates jobs to the cluster resources. This is done while considering various factors, including but not limited to resource pricing, application fault-tolerability, QoS of the applications, resource utilization, and cost.

<u>Autoscaler:</u> If the scheduler finds that there are not enough available resources to allocate to critical jobs, it will interact with the auto-scaler to provide additional resources. The auto-scaler will then decide the type (on-demand, reserved, or rebated) and the amount of resources to be added to the virtual cluster and then provision the new resources through calling the API of the underlying Cloud provider. The autoscaler is also responsible for terminating or migrating containers running on a specific host in order to shut down the host at the end of the host's billing period if the cluster is underutilized. The killed containers will be rescheduled by the scheduler immediately after their termination or be queued and rescheduled later.

<u>Task Launcher:</u> This entity is responsible for launching containers on specific machines and specifying the amount of resources that should be allocated to each container.

<u>Resource Monitor:</u> Monitors the real-time resource consumption such as CPU and memory on each host in the cluster and provides the information to the scheduler to make resource allocation decisions.

<u>Task Monitor:</u> This component is responsible for auditing a running containerized task by recording its resource consumption and monitoring its QoS metrics. This information aides in detecting faults or QoS violations and enables the system to make better scheduling or relocation decisions.

<u>Resource Estimator:</u> This module is used to predict and estimate the amount of resources such as memory and CPU that a container consumes at different points in time. It aims to reduce the framework's reliance on the amount of resources requested by users when submitting their applications. The existence of this component is based on two main assumptions. Firstly, resource requests are usually misestimated, and overestimated, by users. Secondly, the resource consumption of a task is likely to vary over time, with the peak consumption spanning only over a fraction of its lifetime. Both scenarios lead to resources that are reserved but are idle most of the time and hence lead to the cluster being underutilized. By monitoring and estimating the resource consumption of containers, better oversubscription and opportunistic scheduling decisions can be made by the system.

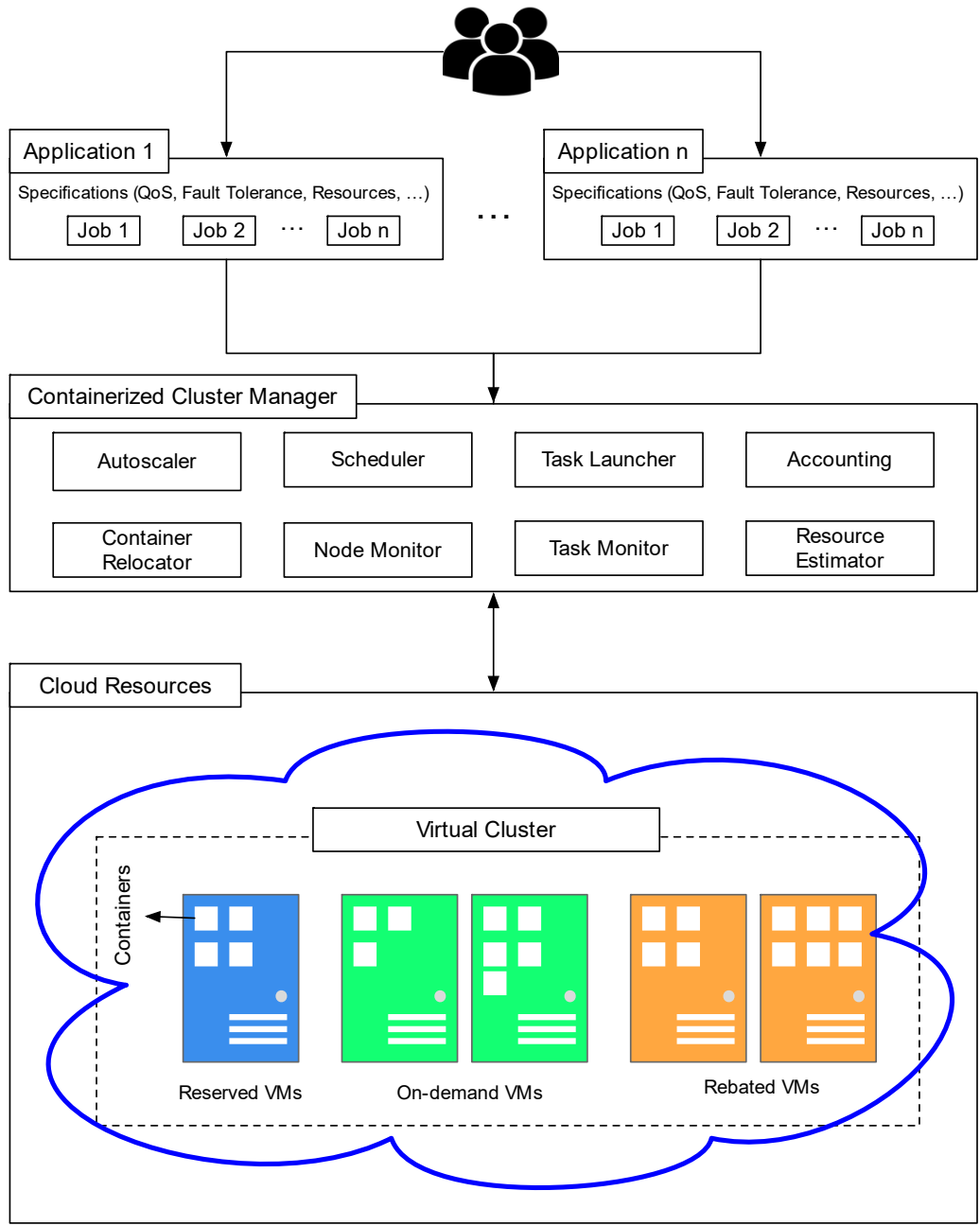

**Figure 1:** A high-level system architectural framework.

*Accounting:* Maintains the actual usage of resources by requests to calculate real-time usage costs. Historical usage information can also be used to improve service allocation decisions.

*Cloud Resources:* Infrastructure as a Service Clouds provide the main compute infrastructure for the deployment of the containerized applications. VMs can be leased on demand and their prices vary based on their capabilities (e.g., amount of CPU and RAM) and their pricing model (e.g., reserved vs. on demand).

# 5. Architectural Elements, Challenges, and Directions
In this section, we discuss various key elements identified in system architecture along with challenges and issues involved in their realization.

## 5.1 Initial Container Placement
Since container migration has to be conducted in a stop and resume manner, its feasibility depends on the nature of the applications. For example, a stateless web service container can be straightforwardly killed and restarted somewhere else; a fault-tolerant data analytics job running in a container can be check-pointed and resume its execution remotely. These containers can be flexibly migrated during runtime. On the other hand, an interactive web service with memory states cannot be migrated without affecting the end users; a time-critical map-reduce job cannot afford the time cost of being interrupted. For these containers, the management platform can only release the resources they use after they completed execution. Therefore, without live migration support, the initial placement algorithm needs to be aware of the application characteristics and as much as possible, avoid the co-location of these two types of containers on the same VM in order to maximize the opportunity in the future for migrating all the containers on it and shut it down. Currently, there is no such consideration in the existing container orchestration platforms as containers are not moveable in these systems. Their target is to simply bin-pack the containers onto as few VMs as possible.

Besides, we believe the placement algorithm in Cloud should also be aware of the heterogeneities of the underlying Cloud resources, including different pricing models, locations, and resource types and sizes. Taking the influence of the pricing to the resource provisioning as an example, a customer-facing application should not be placed on unreliable rebated resources, whose sudden termination will disrupt the end user experience, and should be avoided to be placed on on-demand resources, as it prolongs the running time of expensive on-demand resources. To realize these goals, it is required to filter unqualified resources and propose new resource affinity models to rank the resources when provisioning each framework. The newly proposed policies can be implemented as extensions of the existing filtering and affinity ranking mechanisms of the current platforms or as pluggable schedulers.

## 5.2 Dynamic Rescheduling
The quality of the placement of the containers may degrade as the time passes with workload fluctuations, launchings of containers, and terminations of containers. Therefore, it is necessary to optimize the placement of containers during runtime through migration.

As mentioned earlier, migration is not applicable to all the containers. Thus, the optimization is limited to the VMs whose hosted containers are all moveable. In addition, to fully utilize the acquired VMs, the optimization is conducted at the end of the billing period of each VM. Once the VM is close to the end of the current billing period, the platform judges whether it is possible to terminate all the containers currently running on that VM while not breaking any QoS constraints. If it is feasible, the containers are check-pointed if necessary and killed. The VM is then removed from the virtual cluster and is terminated. The scheduler immediately reacts to the killing of the corresponding containers. If there are free resources, they can be immediately assigned to the killed containers; otherwise, some tasks are queued until other tasks finish and resources are freed.

## 5.3 Dynamic Virtual Cluster Provisioning
Resource contention can happen when some applications ask for a large amount of resources to meet their QoS requirements or urgent tasks are submitted to the virtual cluster. To meet the dynamic resource needs, it is essential to timely provision new resources. On the other hand, the provisioned resources should be just enough and carefully composed regarding pricing models and sizes to minimize the resource cost. Furthermore, the provisioning policy should also be aware of the application nature in order to satisfy the affinity requirements of the applications and as much as possible separate moveable and non-moveable containers.

To integrate dynamic provisioning capabilities into the existing platforms, it is required to modify the architecture of the platform. The new architecture should be able to provide interfaces to monitor the QoS of each application and estimate the resource needs of the application using individually-defined metrics and models.

*5.4 Application QoS Management*
It is not unusual for applications to have specific QoS requirements, support for which is limited in existing systems. For instance, long-running services commonly have to serve a minimum amount of request per time unit or have stringent latency requirements. Batch jobs on the other hand can have a deadline as a time constraint for their execution or may need to be completed as fast as possible. For the first scenario, many systems offer a basic autoscaling mechanism. It monitors the CPU utilization of a service, and if a predefined threshold is exceeded, another instance of the service is launched. This however, is a baseline approach to application autoscaling and integrating more sophisticated approaches to container-based management systems is required. For batch jobs, orchestrating them and assigning them to resources so that their QoS are met is another open research area. For example, mapping tasks to resources so that their makespan is minimized is a useful feature lacking in current open source cluster management systems.

*5.5 Resource Consumption Estimation*
Especially useful when it comes to long-running services, estimating the amount of resources such as memory and CPU that they consume over time can aid in the efficient use of the cluster resources. This mechanism is absent in existing open-source systems, and, to the best of our knowledge, only Google's proprietary platform, Borg [7], has this feature in place. To deal with users overestimating their request for resources when submitting jobs, Borg monitors their resource consumption over time and based on the collected data, predicts their actual resource consumption. Based on this prediction, servers are oversubscribed to run lower-priority tasks that are able utilize the requested but unused resources. Another motivating factor for predicting the resource consumption of long-running services is the fact that it is likely to vary over time, with the peak consumption spanning only over a fraction of its lifetime. If not accounted for, this will also lead to underutilized resources that could be used opportunistically instead.

There are various challenges associated with achieving this goal. Firstly, accurately predicting the resource consumption of jobs is a difficult endeavor. A method capable of incrementally building a model based on data collected over time is essential. Furthermore, such method must be capable of processing streaming data in real time, as this is the nature of resource consumption measurements in a multi-tenant cluster environment. Finally, since the data corresponds to measurements taken in a cloud environment, being able to reflect the dynamicity and performance variability inherent to such platforms is essential. Hence, a method capable of handling concept drift that captures the change in the statistical properties of the collected data is essential. Aside from the analytics and machine learning point of view, successfully using the obtained knowledge to better utilize resources is another challenging area. Integrating the gained knowledge into schedulers and autoscalers must be done while ensuring that the overall system goals such as scalability and job throughput are met while respecting the applications' requirements and QoS goals.

**6. Performance Evaluation – Sample Results**
In this section, we present a prototype software platform that supports the orchestration of containers in clouds and show some of our preliminary experiment results.

*6.1 System Prototype Architecture*
We have implemented a system prototype including the components depicted in Figure 2 by extending the Kubernetes (K8s) platform. The Kubernetes framework is designed to manage containerized workloads on clusters with its basic building block being a pod. A pod encapsulates one or more tightly coupled containers that are co-located on the same machine and share the same set of resources; they also encapsulate storage resources, a network IP, and a set of options that govern how the pod container(s) should run. A pod is designed to run a single instance of an application; in this way multiple pods can be used to scale an application horizontally for example. The amount of CPU, memory, and ephemeral storage a container needs can be specified when creating a pod. This information can then be used by the scheduler to make decisions on pod placement.

Kubernetes is a highly mature system; it stemmed from ten years of experience at Google and is the leading container-based cluster management system with an extensive community-driven support and development base. It provides users with a wide range of options for managing their pods and the way in which they are scheduled, even allowing for pluggable customized schedulers to be easily integrated into the system. We take advantage of this feature to build our platform. Furthermore, as of version 1.10, Kubernetes is capable of supporting clusters of up to 5000 hundred nodes [17], which suits the needs of many organizations nowadays.

A custom scheduler interacts with the Kubernetes API server to continuously monitor the state of pods in the cluster. In particular, it focuses on processing pending pods (i.e., those that need to be scheduled). For each pending pod, a set of suitable resources (i.e., Kubernetes workers) is filtered from the entire cluster pool. One of these resources is then selected and a binding between the pod and the resources is created. This binding leads to Kubernetes running the pod on the chosen node. It is worthwhile noticing that different policies to select the set of suitable nodes and assign a task to one of them are easily pluggable into the system.

The custom scheduler interacts with the custom autoscaler when it determines that more resources are needed to place a pending pod. The autoscaler then decides the number and type of VMs to launch and instructs the Cloud Adapter to create the new instances via the specific IaaS cloud provider API. In the meantime, the unschedulable pod can be left in the scheduling queue in a pending state so that it can be scheduled in a later cycle when the newly provisioned (or recently freed) resources become available. The pod can also be removed from the general scheduling queue so that it can be directly assigned to the newly created node once it is available for use.

The custom scheduler can also make use of the Resource Consumption Estimator when making decisions. For instance, a batch job may be placed in a node in which no all the requested resources are predicted to be used. Different analytical techniques can be plugged into the Resource Consumption Estimator to make predictions on the amount of resources consumed by applications. Currently, a simple statistical technique has been implemented in which the median or average CPU and memory usage is estimated based on historical data. Kubernetes provides its own metric monitoring system as default. That consists of multiple components but the main three components are Heapster as a aggregator, InfluxDB as a time series database and Grafana as a visualizing and alerting solution. Our Resource Consumption Estimator can query to the InfluxDB directly.

*6.2 Scheduling and Autoscaling Policies*

Currently, we have implemented one simple scheduling algorithm and two autoscaling policies for validation purposes.

For each pending pod, a random scheduler first filters all available nodes based on their remaining resource capacity and the amount of resources requested by the pod. These resources are expressed in terms of memory and CPU. Once all nodes with sufficient capacity to execute the pod are identified, a random one is selected, and the pod assigned to it. If there are no nodes that can fulfil this condition, then the scheduler instructs the autoscaler to scale out. The unschedulable pod is left unchanged and an attempt to schedule it again will be made in the next cycle.

The scale out operation will depend on the autoscaler being used. The void autoscaler will simply ignore the request and hence simulates a system without autoscaling capabilities. The simple autoscaler on the other hand, will launch a new instance of a predefined type. The number of instances launched is capped to one every provisioning_interval. The motivation behind this limit is based on the following observation. Unschedulable pods are likely to be found in batches. That is, if there are insufficient resources to deploy one pod, there may be insufficient resources to deploy the next pending pod in the queue. Hence, scaling out requests are likely to be made several times during the same scheduling cycle.

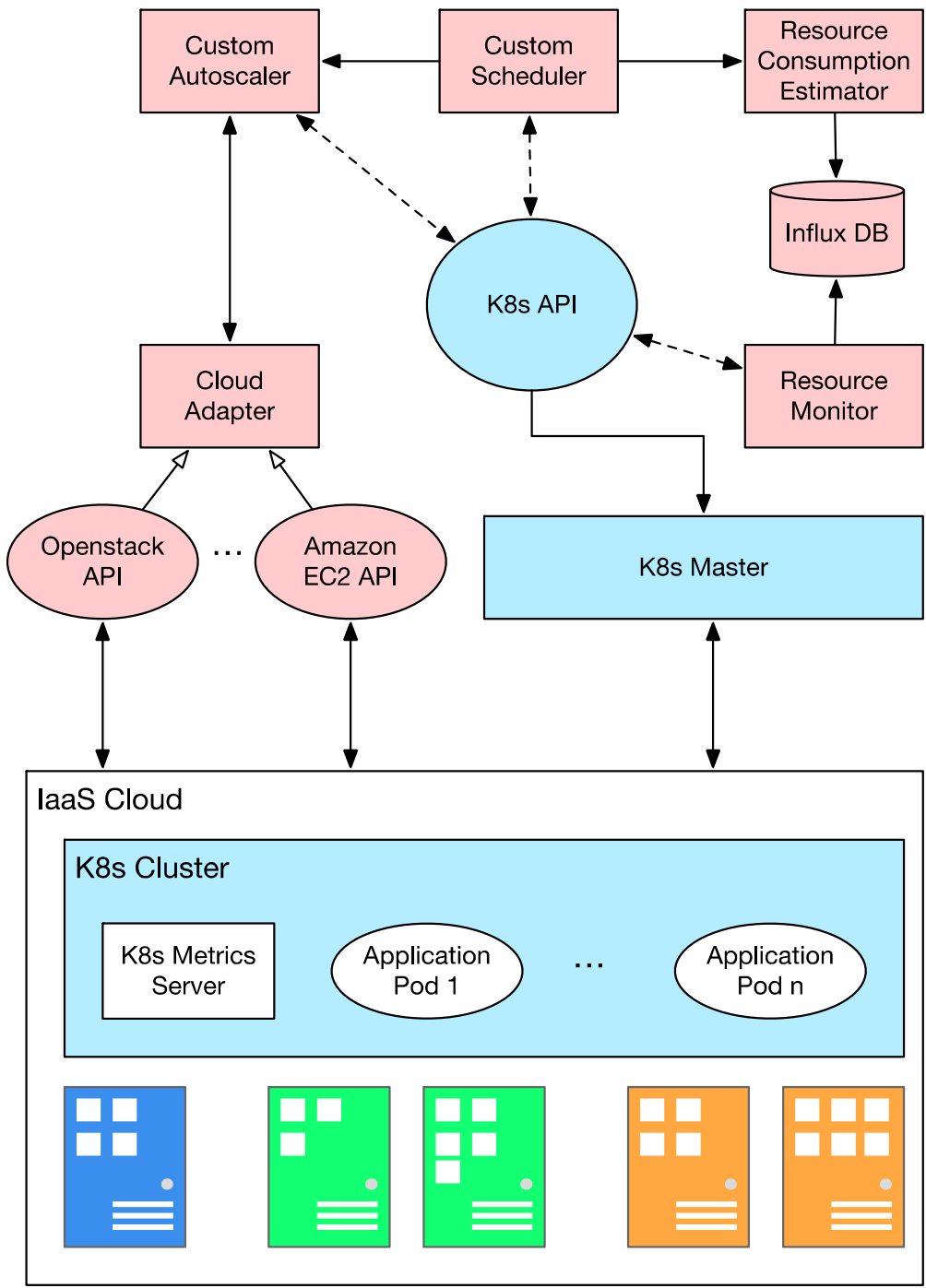

**Figure 2:** System Prototype Architecture.

This may lead to an excessive number of instances being launched which may end up being underutilized as a single one may have sufficed to execute the unschedulable pods. In fact, we set the provisioning_interval based on an estimate of the instance provisioning delay (i.e., the time it takes for the VM to boot and join the K8s cluster) plus a small contingency value. Notice however that this parameter is configurable by users.

**Table 1.** Master and worker nodes VM specifications.

|        | VM Type   | # of vCPUs | RAM  | Operating System |
|--------|-----------|------------|------|------------------|
| Master | m2.medium | 2          | 6 GB | Ubuntu 17.01     |
| Worker | m2.small  | 1          | 4 GB | Ubuntu 17.01     |

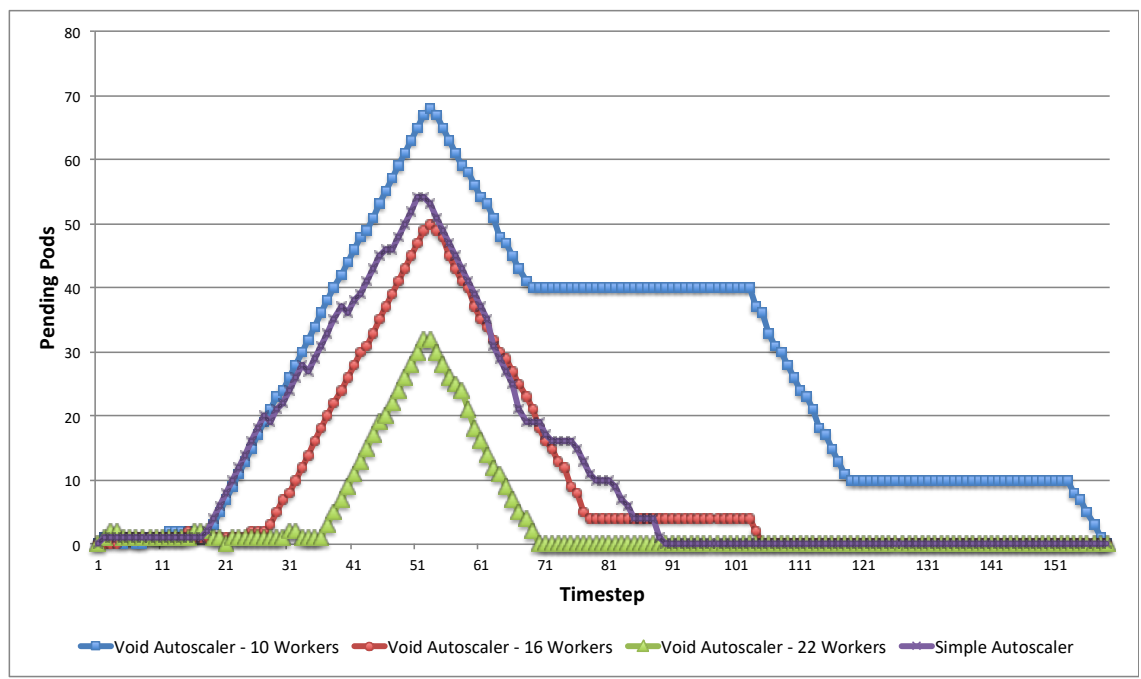

**Figure 3.** Number of pending pods throughout multiple scheduling cycles for different autoscaling scenarios.

*6.3 Testbed*

To demonstrate the potential capabilities of the proposed framework and the benefits of autoscaling, we performed a set of medium-scale validating experiments. For this purpose, we used a workload composed of 100 homogeneous batch jobs, each submitted every 10 seconds to Kubernetes. Each job requests 64 MB of memory and 250 millicores. The jobs do not actually use the requested resources as they only print out a statement and sleep for 1000 seconds (approximately 16 minutes). The purpose of the requests however is to validate the functionality of the scheduler.

The platform is deployed on Nectar [18], an Australian research cloud based on Openstack. The VM specifications used to deploy the Kubernetes master and worker nodes are depicted in Table 1. The custom architectural components were deployed outside Nectar on a MacBook Pro with a 2.9 GHz Intel Core i7 processor and 8 GB of RAM and the Kubernetes version used was 1.10.

*6.4 Results*

The evaluation was performed by comparing 4 different scenarios. The first one, referred to as Simple Autoscaler corresponds to an approach that uses the random scheduler along with the simple autoscaler with an initial cluster size of 10 worker nodes. The three other scenarios all refer to approaches using the random scheduler along with the void autoscaler. They differ from each other in the number of worker nodes in the cluster, which remains static throughout the execution of the workload and ranges from 10 to 22 worker nodes. We refer to these approaches as Void Autoscaler – N workers, where N corresponds to the size of the cluster.

To illustrate the benefits of autoscaling, we evaluate each solution from two perspectives, the scheduling performance and the total infrastructure cost. Figure 3 illustrates the total number of pending pods at different timesteps of the workload execution. Pending pods are those that require scheduling and each monitoring timestep is 20 seconds apart. Clearly, the void autoscaler with 10 nodes takes the longer to schedule all pending pods, followed by the 16-worker node approach. The void autoscaler with 22 nodes is the fastest out of all the approaches in scheduling the pending pods. Finally, the simple autoscaler gradually reduces the number of pending pods as more worker nodes are added to the cluster. Figure 4 shows the number of worker nodes used over time by the simple autoscaler. This approach is able to complete the scheduling the workload soon after the 22-worker approach and before the 16-worker one.

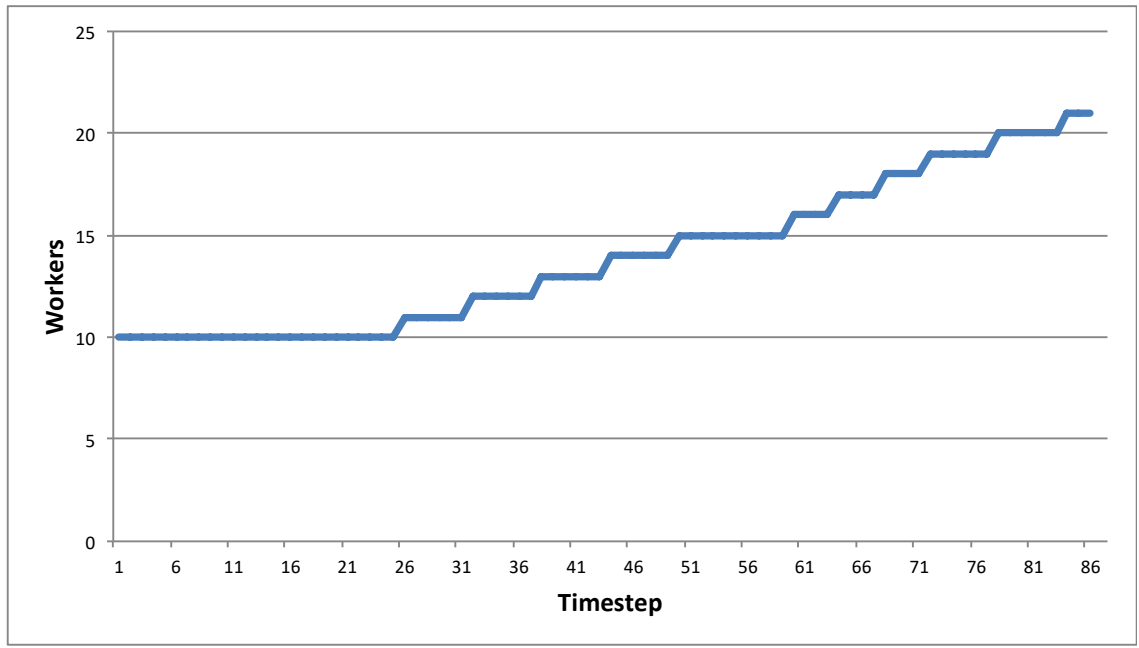

**Figure 4.** Number of worker nodes used by the simple autoscaler at different scheduling cycles.

To demonstrate that despite taking slightly longer to schedule all pods, the simple austoscaler has other benefits, primarily in terms cost, Table 2 depicts various scheduling performance metrics and the total infrastructure cost for each of the solutions. As already mentioned, the 22-worker void autoscaler approach presents the shortest scheduling duration, followed by the simple autoscaler and the 16 and 10-worker void autoscalers respectively. Based on this, the simple autoscaler achieves the second highest pod throughput, followed by the 16 and 10-worker approaches. It is clear that 10 nodes are not sufficient to execute the submitted workload efficiently, with significantly lower throughput measures.

Since Nectar does not charge for the use of resources, we estimate the cost of each approach based on the billing model of existing cloud providers. In particular, we assume a per-minute billing of $0.011 for each worker based on Microsoft Azure's general purpose *B2S* instance type, with any partial use being rounded up to the nearest minute. For the void autoscaler approaches, the number of minutes each worker is billed for is estimated based on the total scheduling duration. For the simple autoscaler, 10 workers are billed for the entire scheduling duration while the remaining 11 are billed from the moment they were launched until the moment when the scheduling of all pods completes. As expected, the simple autoscaler obtains the lowest cost as VMs are only launched when needed. The experiments demonstrate as well that having 22 worker nodes always available only improves the scheduling throughput by 1 pod per minute but leads to higher costs. Also, having a small number of nodes, such as 10 in this case, does

not necessarily reduce the cost, as they must remain active for a longer period of time, inevitably incurring in more billing periods and higher costs.

Table 2. Scheduling performance metrics and cost for different autoscaling scenarios.

| Autoscaler | Average Scheduling Delay (min) | Total Scheduling Duration (min) | Throughput (pods/min) | Cost |
|---|---|---|---|---|
| Void - 22 workers | 1.95 | 23.08 | 4.33 | $0.41 |
| Void - 16 workers | 4.85 | 34.58 | 2.89 | $0.44 |
| Void - 10 workers | 14.03 | 52.71 | 1.89 | $0.42 |
| Simple | 6.13 | 29.59 | 3.37 | $0.32 |

## 7. Summary and Conclusions

This paper presented a vision, an architectural framework, elements, principles, and some preliminary experimental results for Pricing-, Fault-, and QoS-aware Containerized Cluster Management in Clouds. Many opportunities and open challenges exist in this context. Determining the initial placement of containers is one of them; this must be done while considering the characteristics of different applications, cloud providers and pricing models, resource types, geographical location, etc. Another challenge is optimizing the placement of containers at runtime by using rescheduling and migration techniques. The fault-tolerance and QoS requirements of applications must be considered to successfully achieve this objective. Autoscaling the cluster is another important requirement that needs attention;
to meet the dynamic resource needs and to reduce cost it is essential to timely provision new resources when needed and shutdown resources when they are being underutilized. Orchestrating containers so that the applications' QoS constraints are met is another existing challenge. For example, mapping tasks to resources so that their makespan is minimized is a useful feature lacking in current open source cluster management systems. Finally, estimating the amount of resources such as memory and CPU that applications consume over time can aid in the efficient use of the cluster resources. This however is a challenging task that requires further research.

    The evaluation of the new architecture and its elements along with the proposed or new approaches addressing the aforementioned challenges can be fostered using our ContainerCloudsim [16] simulator and the presented empirical platform. For a large-scale evaluation, our simulation toolkit can speed up the evaluation process with various measurements. For a practical proof-of-concept experiment, the empirical evaluation platform can be exploited to see the effectiveness of the proposed methodology in the real world.

**Acknowledgments**
We acknowledge Dr. Chenhao Qu for his contributions towards various ideas presented in this paper. This work supported through a collaborative research agreement between the University of Melbourne and Samsung Electronics (South Korea) as part of the Samsung GRO (Global Research Outreach) program. The work is carried out within the Melbourne CLOUDS Laboratory at the University of Melbourne.